\documentclass[conference]{IEEEtran}
\usepackage[]{times}
\usepackage{epsfig}
\usepackage{subfigure}
\usepackage{amsmath}
\usepackage{amssymb}
\usepackage{graphicx}
\usepackage{multirow}
\usepackage{algorithm}
\usepackage{mathrsfs}
\usepackage{url}
\usepackage[usenames]{color}
\newcommand{\sgn}{\text{sgn}}

\newtheorem{defn} {Definition}
\newtheorem{cor} {Corollary}
\newtheorem{te}{Theorem}
\newtheorem{lem}{Lemma}

\newcommand{\mrm}{\mathrm}

\begin{document}
\title{Decimation-Enhanced Finite Alphabet Iterative Decoders for LDPC codes on the BSC}
\author{\IEEEauthorblockN{Shiva Kumar Planjery, Bane Vasic}
\IEEEauthorblockA{Dept. of Electrical and Computer Eng.\\
University of Arizona\\
Tucson, AZ 85721, U.S.A.\\
Email: \{shivap,vasic\}@ece.arizona.edu}
\and
\IEEEauthorblockN{David Declercq}
\IEEEauthorblockA{ETIS\\
ENSEA/UCP/CNRS UMR 8051\\
95014 Cergy-Pontoise, France\\
Email: declercq@ensea.fr}
}
\maketitle
\begin{abstract}
Finite alphabet iterative decoders (FAID) with multilevel messages that can surpass BP in the error floor region for LDPC codes on the BSC were previously proposed in \cite{planjery}. In this paper, we propose decimation-enhanced decoders. The technique of decimation which is incorporated into the message update rule, involves fixing certain bits of the code to a particular value. Under appropriately chosen rules, decimation can significantly reduce the number of iterations required to correct a fixed number of errors, while maintaining the good performance of the original decoder in the error floor region. At the same time, the algorithm is much more amenable to analysis. We shall provide a simple decimation scheme for a particularly good 7-level FAID for column-weight three codes on the BSC, that helps to correct a fixed number of errors in fewer iterations, and provide insights into the analysis of the decoder. We shall also examine the conditions under which the decimation-enhanced 7-level FAID performs at least as good as the 7-level FAID.
\end{abstract}
%\begin{center}
%\textbf{\small Index Terms}
%\end{center}
%{\small Low-density parity-check codes, trapping set, iterative decoding algorithm}
%
\section{Introduction}\label{sect_Intro}
%both from a theoretical as well as a practical standpoint (removed from intro)
%and deriving conditions for guaranteed error-correction capability  (removed from intro first para)
%and they provide insights into design of LDPC codes with good performance in the waterfall region.(removed from intro first para)
%The decimation procedure could be deterministic or randomized. %and EXIT charts \cite{brink} by ten Brink et. al.,
The design and analysis of message-passing (MP) algorithms for low-density parity-check (LDPC) \cite{gallager} codes have recieved much attention over the last decade. Techniques such as density evolution \cite{richardson} by Richardson and Urbanke, have been proposed for asymptotic analysis of MP decoders on LDPC code ensembles. For finite-length analysis of codes with fixed number of iterations, methods such as the use of computation trees by Wiberg \cite{wiberg}, pseudocodeword analysis by Kelly and Sridhara \cite{kelly}, and graph-cover decoding analysis by Vontobel and Koetter \cite{vontobel}, have been proposed. The characterization of the error floor phenomenon of MP algorithms has also been well investigated using the notion of stopping sets for the binary erasure channel (BEC)\cite{Di} by Di et. al., and using notions of trapping sets by Richardson \cite{richardsontrap} and instantons by Chernyak et. al. \cite{chernyak} for other general channels. Burshtein and Miller proposed the technique of using expander arguments for MP for proving that code ensembles can correct a linear fraction of errors \cite{Burshtein}.  

Inspite of the aforementioned techniques proposed for finite-length analysis, the problem of analyzing a particular MP algorithm for a fixed number of iterations still remains a challenge. This is because the dynamics of MP gets too complex beyond a certain number of iterations, and there is exponential growth in the number of nodes with number of iterations in the computation trees of the codes. Although Burshtein and Miller's method of using expander arguments which allows for use of large number of iterations, provides bounds of great theoretical value, they are practically less significant. Moreover, for the Binary Symmetric channel (BSC), the problem of correcting a fixed number of errors assumes greater importance as it determines the slope of the error floor in the performance of the decoder \cite{milos}. Therefore, it would be desirable to have an MP decoder that is able to correct a fixed number of errors within fewest possible iterations, and for which we will be able to provide performance guarantees in terms of guaranteed correction capability. Even from a practical standpoint, this would be an attractive feature with many present-day applications requiring much higher decoding speeds and much lower target frame error rates. 

Recently we proposed a new class of finite alphabet iterative decoders (FAID) for LDPC codes on the BSC coined as \textit{multilevel decoders} in \cite{planjery} and showed that these decoders have potential to surpass belief propagation (BP) in the error floor region with much lower complexity. These decoders were derived by identifying potentially harmful subgraphs that could be trapping sets present in any finite-length code and designing to correct error patterns on these subgraphs in an isolated manner. Although the numerical results in \cite{planjery} demonstrated the efficacy of these decoders, providing provable statements in terms of guaranteed error correction capability still remains a difficult task since the convergence of the decoder for an error pattern in a trapping set is heavily influenced by the neighborhood of the trapping set in a non-trivial manner. This was also identified by Declercq et. al. in \cite{Davidbrest}, where subgraphs induced by codewords were used in the decoder design.  

In this paper, we propose {\it decimation-enhanced finite alphabet iterative decoders} for LDPC codes on the BSC. Decimation, a method originally developed for solving constraint satisfaction problems in statistical physics, involves guessing the values of certain variables and fixing them to these values while continuing to estimate the remaining variables. In \cite{montonari}, Montanari et. al. analyzed a BP-guided randomized decimation procedure that estimates the variables in the \textit{k}-SAT problem. Dimakis and Wainwright used a similar notion in the form of facet guessing for linear programming (LP) based decoding in \cite{dimakis}, and Chertkov proposed a bit-guessing algorithm in order to reduce error floors of LDPC codes under LP decoding \cite{chertkov}. In contrast, we propose a simple decimation procedure in this paper that serves as a guide to help the multilevel FAID algorithm to coverge faster on a small number of errors. Our main insight is that the role of decimation should not necessarily be to correct errors, but to ensure that more variable nodes in the graph that initially receive right values from the channel are shielded from the errorneous messages emanating from the error nodes by decimating those correct variable nodes. 

The rest of the paper is organized as follows. Section \ref{sect_Prelim} provides preliminaries. In Section \ref{sect_faid}, we provide a detailed description of the decimation-aided FAID algorithm. Finally in Section \ref{sect_analysis}, we provide some theoretical as well as numerical results and conclude with a discussion.

\section{Preliminaries}\label{sect_Prelim}
Let $G=(V\cup C,E)$ denote the Tanner graph of an ($n$,$m$) binary LDPC code $\cal{C}$ with the set of variable nodes $V=\{v_1,\cdots,v_n\}$ and set of check nodes $C=\{c_1,\cdots,c_m\}$. $E$ is the set of edges in $G$. A code $\cal{C}$ is said to be $d_v$-left-regular if all variable nodes in $V$ of graph $G$ have the same degree $d_v$. The degree of a node is the number of its neighbors.

Let $\mathbf{r}=(r_1,r_2\ldots,r_n)$  be the input to the decoder from the BSC. A trapping set $\mathbf{T}(\mathbf{r})$ is a non-empty set of variable nodes in $G$ that are not eventually corrected by the decoder \cite{richardsontrap}. Note that during analysis of decoders in this paper, it is implicitly assumed that the all-zero codeword is transmitted. This is a valid assumption since we consider only symmetric decoders, as explained in \cite{richardson}. 

A multilevel FAID $\mathscr{F}$, as defined in \cite{planjery}, is a 4-tuple given by $\mathscr{F}=(\mathcal{M},\mathcal{Y},\Phi_v,\Phi_c)$. The messages are levels confined to a message alphabet $\cal{M}$ which is defined as $\mathcal{M}=\{ 0,\pm L_i \ :1\leq i\leq M  \}$, where $L_i\in\mathbb{R^{+}}$ and $L_i>L_j$ for any $i>j$. The set $\mathcal{Y}$ denotes the set of possible {\it channel values}. For the case of BSC, $\mathcal{Y}$ is defined as $\mathcal{Y}=\{\pm \mrm{C}\}$, and for each variable node $v_i$ in $G$, the channel value $y_i\in \cal{Y}$ is determined by $y_i=(-1)^{r_i}\mrm{C}$, i.e., we use the mapping $0\rightarrow \mrm{C}$ and $1\rightarrow -\mrm{C}$.

$\Phi_v:\mathcal{Y} \times \mathcal{M}^{d_v-1} \to \mathcal{M} $ is the update rule used at a variable node with degree $d_v$. The map $\Phi_v$ can be described in closed form as a linear or non-linear threshold function, or simply as a look-up table (LUT). For this paper, we shall use the LUT form. The LUT of $\Phi_v$ for a particularly good 7-level FAID that we shall be using throughout this paper is given in Table \ref{LUT244325} for $y_i=+\mrm{C}$ (for $y_i=-\mrm{C}$, the LUT can be obtained by symmetry). 
 
$\Phi_c: \mathcal{M}^{d_c-1} \to \mathcal{M} $ is the update function used at a check node with degree $d_c$. The function is given by
\begin{displaymath}
\Phi_c(m_1,\ldots,m_{d_c-1}) = \left(\prod_{j=1}^{d_c-1}\sgn(m_j)\right) \min_{j \in \{1,\ldots,d_c-1\}}(|m_j|)
\end{displaymath}
where $\sgn$ denotes the standard signum function.

\begin{table}[tp]
	\centering
  \caption{LUT of $\Phi_v$ used for the 7-level decoder for $y_i=+\mrm{C}$}
	  
		\begin{tabular}{|c|c|c|c|c|c|c|c|}
		\hline $m_1\backslash m_2$ & -$L_3$ & -$L_2$ & -$L_1$ & 0  & $L_1$ & $L_2$ & $L_3$\\
		\hline 	-$L_3$	 &	-$L_3$	&	-$L_3$	&	-$L_2$	&	-$L_1$	&	-$L_1$	&	-$L_1$	&	$L_1$\\
		\hline	-$L_2$	 &	-$L_3$	&	-$L_1$	&	-$L_1$	&	0	&	$L_1$	&	$L_1$	&	$L_3$\\
		\hline	-$L_1$	&	-$L_2$	&	-$L_1$	&	0	&	0	&	$L_1$	&	$L_2$	&	$L_3$\\
		\hline	0				&	-$L_1$	&	0	&	0	&	$L_1$	&	$L_2$	&	$L_3$	&	$L_3$\\
		\hline	$L_1$		&	-$L_1$	&	$L_1$	&	$L_1$	&	$L_2$	&	$L_2$	&	$L_3$	&	$L_3$\\
		\hline	$L_2$		&	-$L_1$	&	$L_1$	&	$L_2$	&	$L_3$	&	$L_3$	&	$L_3$	&	$L_3$\\
		\hline	$L_3$		&	$L_1$	&	$L_3$	&	$L_3$	&	$L_3$	&	$L_3$	&	$L_3$	&	$L_3$\\
		\hline
		\end{tabular}
	
\label{LUT244325}
\end{table}
%wherein the update rules for $\Phi_v$ were derived based on analyzing the decoding of error patterns on potentially harmful subgraphs assuming that the rest of the graph outside the subgraph is error free.
The important concept of \textit{isolation assumption} was also introduced in \cite{planjery} and here we remind the reader on the intution behind it. The isolation assumption provides conditions on the neighborhood of the subgraph, such that the messages entering into the subgraph from outside are not affected by the messages being propagated within the subgraph for a certain a number of iterations. Consequently, for a certain number of iterations, decoding on the subgraph can be carried out in an isolated manner without explicit knowledge of its neighborhood.

%{\it Remark: } In reality however, the neighborhood of the subgraph will eventually play a role and affect the decoder convergence, which complicates the decoding analysis. %This further motivates the incorporation of decimation into the decoding algorithm which we shall elaborate on in the next section. 
 
\section{Decimation-enhanced faid algorithm}\label{sect_faid}
We first provide some basic definitions and notations before we formally introduce the class of decimation-enhanced finite alphabet iterative decoders.  

Let $\mathcal{N}(u)$ denote the set of neighbors of a node $u$ in the graph $G$ and let $\mathcal{N}(U)$ denote the set of neighbors of all $u\in U$. Let $m_k(v_i,c_j)$ denote the message being passed from a variable node $v_i$ to the check node $c_j$, in the $k^{th}$ iteration, and let $m_k(c_j,v_i)$ be defined similarly. Let $m_k(v_i,\mathcal{N}(v_i))$ denote the set of outgoing messages from $v_i$ to all its neighbors in the $k^{th}$ iteration, and let $m_k(c_j,\mathcal{N}(v_i))$ be defined similarly. Let $b_i^{k}$ denote the bit associated to a variable node $v_i\in V$ that is decided by the iterative decoder at the end of the $k^{th}$ iteration.

\begin{defn}
A variable node $v_i$ is said to be {\it decimated} at the end of $l^{th}$ iteration if $b_i^{k}=b_i^{\ast}$ $\forall k\geq l$. Then 
$m_k(v_i,\mathcal{N}(v_i))=\{(-1)^{b_i^{\ast}}L_M\}$, $\forall k\geq l$ irrespective of its incoming messages $m_k(\mathcal{N}(v_i),v_i)$, i.e., $v_i$ will always send strongest possible messages. 
\end{defn}
The process of decimation at the end of some $l^{th}$ iteration is carried out by the iterative decoder using a \textit{decimation rule} $\beta:\mathcal{Y} \times \mathcal{M}^{d_v} \to \{-1,0,1\}$ that is a function of the incoming messages and the channel value in the $l^{th}$ iteration. For convenience, with some abuse of notation, let $\beta_i$ denote the output of function $\beta$ determined at node $v_i$. If $\beta_i=0$, then the node is not decimated. If $\beta_i= 1$, then $b_i^{\ast}=0$, and if $\beta_i=-1$, then $b_i^{\ast}=1$.

{\it Remark:} In this paper, we only consider decoders that use a single decimation rule $\beta$ but the same rule may be used in different iterations. Hence, $\beta$ is not iteration dependent, and whenever we refer to $\beta_i$, it implies the output of $\beta$ for node $v_i$ at the end of some $l^{th}$ iteration. Note that the function $\beta$ is symmetric.

A decimation-enhanced multilevel FAID $\mathscr{F}^{D}$ is defined as a 4-tuple given by $\mathscr{F}^D=(\mathcal{M},\mathcal{Y},\Phi_v^{D},\Phi_c)$, where 
$\mathcal{M}$, $\mathcal{Y}$, $\Phi_c$ are the same maps defined for a multilevel FAID. The map $\Phi_v^{D}:\mathcal{Y} \times \mathcal{M}^{d_v-1} \times \{0,1\} \to \mathcal{M}$ is similar to $\Phi_v$ of the decoder $\mathscr{F}$ except that it uses the output of $\beta$ in some $l^{th}$, $\beta_i$, as an additional argument in the function. For the sake of simple exposition, we shall define $\Phi_v^{D}$ for the case of column-weight three codes and 7 levels. Let $m_1$ and $m_2$ denote incoming messages to a node $v_i\in V$ in the $k^{th}$ iteration. Then $\Phi_v^{D}$ is defined as 
\begin{displaymath}
\Phi_v^{D}(m_1,m_2,y_i,\beta_i)=\Bigg\{ \begin{tabular}{ll}
 $\Phi_v(m_1,m_2,y_i)$,&$\beta_i=0$ \\
$\beta_i L_3$,&$\beta_i=\pm 1$ \\
\end{tabular}
\end{displaymath}

We now provide a simple example to illustrate the potential benefits of decimation, and then we will describe the basic decimation scheme used in this paper. 

\subsection{Motivating example}

\begin{figure}[tp]
\begin{center}
\includegraphics[angle=0, width=1.6in]{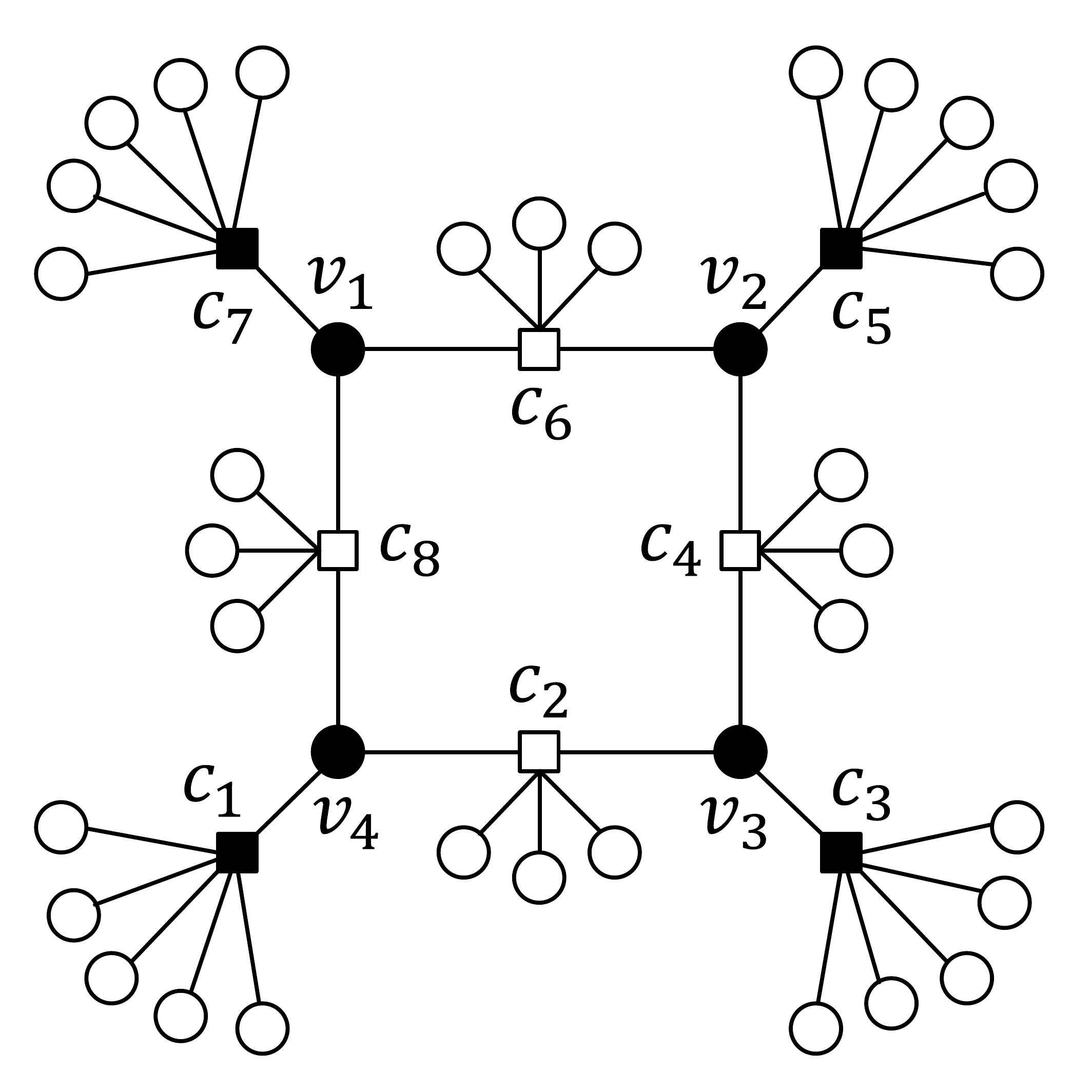}
\caption{Subgraph induced by the 4-error pattern which forms an 8-cycle}
\label{example}
\end{center}
\vspace{-0.2in}
\end{figure}

Consider a particular 4-error configuration on a Tanner graph $G$, whose induced subgraph forms an 8-cycle as shown in Fig. \ref{example}. In the figure, black circles represent the variable nodes initially in error, whereas white circles represent the intially correct nodes that are in the neighborhood of the 4-error pattern. The black and white squares denote the degree one and degree two checks in the induced subgraph respectively. 

Let $V'$=$\{v_1,v_2,v_3,v_4\}$ denote the set of variable nodes initially in error. Let $C^1$=$\{c_1,c_3,c_5,c_7\}$ denote the set of degree one checks and $C^2$=$\{c_2,c_4,c_6,c_8\}$ denote the set of degree two checks. We shall now examine the behavior of MP on this particular error configuration from the context of multilevel FAID algorithms without any assumptions on its neighborhood. Messages with a positive sign will be referred to as \textit{good} messages, and messages with a negative sign will be referred to as \textit{bad} messages (under all-zero codeword assumption). Also a weakly good or bad message refers to $\pm L_1$, and a strongly good or bad message refers to $ \pm L_i$ where $L_i>L_1$. In the first iteration, for all $v_i\in V'$, $m_1(v_i,\mathcal{N}(v_i))$ will be weakly bad, and for all $v_j\in \mathcal{N}(C^1\cup C^2)\backslash V'$, $m_1(v_j,\mathcal{N}(v_j)\cap (C^1 \cup C^2))$ entering into the subgraph will be weakly good messages. In the second iteration, for all $v_i\in V'$, $m_2(v_i,\mathcal{N}(v_i)\cap C^2)$ will be either weakly good or weakly bad depending on the $\Phi_v$ (such as Table \ref{LUT244325}), but $m_2(v_i,\mathcal{N}(v_i)\cap C^1)$ which are messages sent to checks in $C^1$, will be strongly bad. As a result, variable nodes $v_i\in \mathcal{N}(C^1)\backslash V'$ will start receiving strongly bad messages. If the decoder does not converge within the next few iterations, then these bad messages become more stongly bad and can subsequently spread further to other nodes in the graph depending how dense the neighborhood is. Eventually too many nodes get corrupted by the bad messages being propagated in the graph causing the decoder to fail.

{\it Remarks:} 1) At the $k^{th}$ iteration, there may have been many variable nodes $v_i$ such that $v_i\notin \mathcal{N}(C^1\cup C^2)$, whose incoming messages already converged to the right value in some $k'<k$ iteration, but eventually these nodes became corrupted by the bad messages flowing out of the subgraph.
2) If certain variable nodes initially correct in the neighborhood of the subgraph induced from the error pattern, are isolated from the bad messages possibly through decimation, then the decoder is more likely to converge. This is precisely where the role of decimation becomes important.

\subsection{Basic scheme for the decimation-enhanced FAID algorithm} 

We propose a scheme that uses successive decimations for a certain number of iterations. Let the number of successive decimations be $N_d$. The skeleton of the algorithm is given below. Note that for this proposed scheme, decimation starts at the end of the third iteration (reasons for which we will explain later), and  after each decimation, the decoder is restarted.  

\begin{algorithm}
\caption{Decimation-enhanced FAID algorithm}
\begin{itemize}
\item[1)] Initialize $\beta_i=0$ $\forall v_i\in V$.
\item[2)] Run the decoder using update maps $\Phi_v$ and $\Phi_c$ already defined for the 7-level FAID decoder. 
\item[3)] At the end of third iteration, perform decimation using the rule $\beta$ for every node $v_i\in V$.
\item[4)]Restart the decoder by setting all the messages to zero. 
\item[5)] Repeat step 3) for nodes whose $\beta_i=0$, followed by 4) until $N_d$ decimations have been carried out.   
\item[6)] Run the decoder for the remainder of iterations using maps $\Phi_v^{D}$ and $\Phi_c$. 
\end{itemize}
\end{algorithm}
Note that restarting the decoder implies, that the decimated nodes $v_i$ will send $\beta_i L_3$ and the non-decimated nodes $v_j$ will send $\Phi_v(0,0,y_j)$.
 
{\it Remarks:} 1) The reasons for choice of three iterations to start the decimation are as follows. First, there should exist messages in the graph that have a magnitude of $L_3$ in order to have a reliable decimation process. Secondly and more importantly, decimating only after three iterations makes the algorithm much more amenable to analysis. It becomes possible to analyze whether a particular node will be decimated or not and derive explicit conditions on the graph under which the nodes in error will not get decimated to the wrong value for a particular configuration. We shall in fact derive such conditions for the previous example. 2) Restarting the decoder after each decimation simplifies analysis.

\subsection{Design of decimation rule $\beta$}
The design of the rule $\beta$ can be considered as a procedure of selecting the sets of incoming messages to node $v_i$ for which $\beta_i=\pm 1$. We would like to do the selection with particular focus on correcting small number of errors typically associated with trapping sets in the error floor region. Referring back to the previous example, a good decimation rule would be one where $\beta_j$ for most or all nodes $v_j\in \mathcal{N}(C^1\cup C^2)\backslash V'$ is 1  and $\beta_i$ for nodes $v_i \in V'$ is 0 or 1, at the end of all decimations. We will now describe a good decimation rule selected for the particular 7-level FAID whose LUT is shown in Table \ref{LUT244325}. Before we describe the rule, we highlight two main points to be considered during the selection.

Firstly, while considering a particular set of incoming messages, the magnitudes of the incoming messages should play a role in the selection. 

Secondly, the inherent structure of the particular $\Phi_v$ used in the decoder must be taken into consideration during selection. For this, we need to look at what outgoing messages a variable node would send for that particular set of incoming messages, and then decide if this set is good to select for decimation. For example, if a variable node $v_i$ whose channel value is $\mrm{+C}$, receives $-L_2$,$-L_3$,$-L_2$, this set might seem to be a possible candidate (to decimate $v_i$ to 1). However, the outgoing messages will be $-L_3$,$-L_1$,$-L_3$, which perhaps indicates that this may not be a reliable node to decimate since all outgoing messages are not $-L_3$ or even $-L_2$. 

Table \ref{decimatetable} shows all possible distinct sets of incoming messages with $y_i=\mrm{+C}$, for which $\beta_i=1$. Using the symmetry of $\beta$, we can derive the sets of incoming messages with $y_i=-\mrm{C}$, for which $\beta_i=-1$. Note that an important condition that was used in defining the rule $\beta$, was that there must be a strict majority of signs of messages between all the messages coming to a node $v_i$ and channel value $y_i$, and the majority sign must match with the sign of $y_i$, in order for $v_i$ to be decimated. 
%
%\noindent \underline{{\bf Conditions needed to be satisfied for decimation:}}\\
%\noindent 1. There must be strict majority of signs between $m_1$, $m_2$, $m_3$, and channel value $y_i$. In addition, the majority sign must match with the sign of the channel value $y_i$.\\
%\noindent 2. The magnitude of the message with the minority sign must not be greater than $L_1$.\\
%\noindent 3. If the magnitude of at least one incoming message is $L_3$, then the set of outgoing messages should have two messages with magnitude of $L_3$, and one message with magnitude greater than or equal to $L_1$. \\
%\noindent 4. If none of the incoming messages have a magnitude of $L_3$, then all incoming messages must have same sign. In addition two of them must have magnitude of $L_2$ and the third should have a magnitude greater than or equal to $L_1$.

\section{Analysis of decimation-enhanced faid algorithms}\label{sect_analysis}
\subsection{Theoretical results}
We first state the following lemma which is obtained due to the conditions used for decimation.
\begin{lem}
The decimation-enhanced FAID algorithm will never decimate a node initally correct to a wrong value, and a node initially wrong to a correct value.
\end{lem}
\IEEEproof 
By virtue of $\beta$ that requires strict majority of signs of messages between incoming messages and $y_i$, and the majority sign matching sign of $y_i$. 
\endIEEEproof

{\it Remark:} This simplifies the analysis as we need to only be concerned about decimation of nodes that are initially in error. At the same time, note that decimation alone can never correct errors.
\begin{cor}
As a consequence of Lemma 1, the only necessary condition for success of a decimation-enhanced multilevel decoder is that a node initially in error must not be decimated.
\end{cor} 

\begin{table}[tp]
\begin{center}
\caption{Sets of incoming messages with $y_i=\mrm{+C}$ for which $\beta_i=1$}

\begin{tabular}{ccc}
\begin{tabular}{|c|c|c|}
\hline $m_1$ & $m_2$ & $m_3$\\
\hline	$L_3$	&	$L_3$	&	$L_3$\\
\hline	$L_3$	&	$L_3$	&	$L_2$\\
\hline	$L_3$	&	$L_3$	&	$L_1$\\
\hline	$L_3$	&	$L_3$	&	0\\
\hline	$L_3$	&	$L_3$	&	-$L_1$\\
\hline					
\end{tabular} &
\begin{tabular}{|c|c|c|}
\hline $m_1$ & $m_2$ & $m_3$\\
\hline	$L_3$	&	$L_2$	&	$L_2$\\
\hline	$L_3$	&	$L_2$	&	$L_1$\\
\hline	$L_3$	&	$L_2$	&	0\\
\hline	$L_3$	&	$L_2$	&	-$L_1$\\
\hline	$L_3$	&	$L_1$	&	$L_1$\\
\hline					
\end{tabular} &
\begin{tabular}{|c|c|c|}
\hline $m_1$ & $m_2$ & $m_3$\\
\hline	$L_3$	&	$L_1$	&	0\\
\hline	$L_3$	&	$L_1$	&	-$L_1$\\
\hline	$L_3$	&	0	&	0\\
\hline	$L_2$	&	$L_2$	&	$L_2$\\
\hline	$L_2$	&	$L_2$	&	$L_1$\\
\hline					
\end{tabular}
\end{tabular}
\vspace{-0.2in}\label{decimatetable}
\end{center}
\end{table}

\begin{lem}
Given an error pattern, if no node initially in error gets decimated at the end of third iteration, then a node initially in error will never get decimated in the susbsequent iterations.
\end{lem}
\IEEEproof
(Details omitted) By restarting the decoder after each decimation, and by virtue of $\beta$, a node $v_i$ initially in error will not receive the required messages for $\beta_i=-1$.
\endIEEEproof

Now for a particular error configuration in a graph $G$, we can analyze the conditions under which a node that is initially in error is decimated at the end of third iteration and then use Lemma 2.  We can then place conditions on the graph such that the node in error is never decimated. To show this, we revert back to the example of the 4-error configuration whose induced subgraph forms an 8-cycle, and provide such conditions in the following theorem. Note that the proof will involve using Tables \ref{LUT244325} and \ref{decimatetable} and also the same notations previously defined in the example.  

\begin{te}
Consider the 4-error pattern contained in a graph $G$, whose induced subgraph forms an 8-cycle. Also consider the decimation-enhanced 7-level FAID for decoding on this error pattern. If the graph $G$ has girth-8, and no three check nodes of the 8-cycle share a common variable node, then the nodes initially in error will not be decimated by this decoder in any iteration.  
\end{te}
\IEEEproof 
Firstly note that by virtue of $\Phi_v$ of the 7-level FAID (Table \ref{LUT244325}), the highest magnitude of a message that any node $v_i\in V$ can send is $L_1$ in the first iteration and $L_2$ in the second iteration.
Since a node $v_j\in \mathcal{N}(C^1\cup C^2)\backslash V'$ can be connected to atmost two checks in subgraph, the node $v_j$ in the worst case recieves two $-L_1$ messages from checks in $C^1\cup C^2$ and $L_1$ from outside at the end of first iteration. Node $v_i\in V'$ will also receive two $-L_1$ messages from check nodes in $C^2$ and $L_1$ from $c_k\in C^1\cap\mathcal{N}(v_i)$. At the end of the second iteration, the node $v_i\in V'$ will once again receive two $-L_1$ from checks in $C^2$, and $L_1$ from $c_k\in C^1$. This means that node $v_i$ will receive two $-L_1$ messages once gain from checks in $C^2$ at the end of third iteration. In order for it to be decimated, from Table \ref{decimatetable}, it must receive $-L_3$ from $c_k\in C^1\cap\mathcal{N}(v_i)$. This means that the node $v_j$ in the worst case has to receive at least one $-L_3$ at the end of the second iteration, but this is not possible by virtue of $\Phi_v$ in the second iteration. Hence, a node initially in error can not get decimated at the end of third iteration and using Lemma 2, will never get decimated. 
\endIEEEproof

{\it Remarks: }  Note that the above condition is easily satisfied in most practical codes. This implies that on most practical codes, 4 errors on an 8-cycle will not be decimated. 

Similarly, we can analyze under what conditions a node initially correct is decimated. For example, we may be able to derive conditions on the neighbors of the 4-error configuration such that they get decimated. In this manner, we can link analytical results of decimation to guaranteed error-correction capability. 

Given that nodes initially in error are not decimated, it would be interesting to know whether the decimation-enhanced FAID can perform as good as multilevel FAID. In other words, given the necessary condition is satisfied, if a 7-level decoder corrects a particular $K$-error pattern, does the decimation-enhanced decoder also correct the pattern. Intuitively, this may appear to be true since only nodes initially correct are decimated and they continuously send strong correct messages for the entire decoding process. However, it is true only under certain conditions given in the following theorem. 

\begin{te}
Let there be a $K$-error pattern on a code $\mathcal{C}$ with a Tanner graph $G$, whose graph satisfies conditions such that nodes initally in error are not decimated under the decimation-enhanced FAID. Assume that the multilevel FAID corrects this error pattern in $I$ iterations. If under the decimation-enhanced FAID algorithm, the decimation of a particular node does not lead to a bad message being sent from one of its adjacent check nodes in any iteration, then the decimation-enhanced FAID algorithm is also guaranteed to correct the pattern in atmost $I$ iterations, 
\end{te}
\IEEEproof 
Due to page constraints, proof is omitted. But the idea of the proof involves analyzing how messages flowing along the edges of the computation tree towards the root under the FAID algorithm, are affected by using the decimation-enhanced FAID algorithm.        
\endIEEEproof

{\it Remark: } Although Theorem 2 includes a specific condition on the decimation (under the assumption that nodes initially in error are not decimated), such a condition typically occurs for larger errors and decoding with much larger number of iterations. Therefore on small number of errors in trapping sets with a smartly chosen decimation rule, if nodes initially in error are not decimated, then decimation-enhanced FAID will most likely correct the pattern. 

\subsection{Numerical Results and Discussion}
In this subsection, we present numerical results on the well-known (155,93) Tanner code in order to evaluate the performance of the decimation-enhanced FAID. The frame-error rate curves for various decoders are shown in Fig. 2. The decimation-enhanced FAID was run using $N_d$ = 4 and all
decoders used a maximum of 100 iterations. Note that the decimation-enhanced FAID was designed primarily to correct a fixed number of errors (in this case 5 errors) in fewer iterations compared to 7-level FAID. On the Tanner code, with $N_d$ = 1, the decimation-enhanced FAID corrects all 5
errors within 10 iterations (after decimation) whereas the 7-level FAID requires 15 iterations. At the same time, we see that decimation-enhanced FAID performs just as good as the 7-level FAID (which was known beforehand to surpass BP), if not better.

We conclude by mentioning that our main goal was to provide a simple decimation scheme for multilvel FAID decoders that allows us to analyze their behaviour while maintaining good performance. From the theoretical analysis, we see that the role of decimation is important not just in improving the decoder performance or reducing the decoder speed but more so in terms of increasing the feasibility to obtain provable statements on the performance MP decoders such as FAID that are known to be empirically good. We finally remark that with more sophisticated versions of decimation such as use of multiple decimation rules, it might be possible to obtain an even more significant performance improvement.

\begin{figure}[tp]
\begin{center}
\includegraphics[angle=0, width=3.3in]{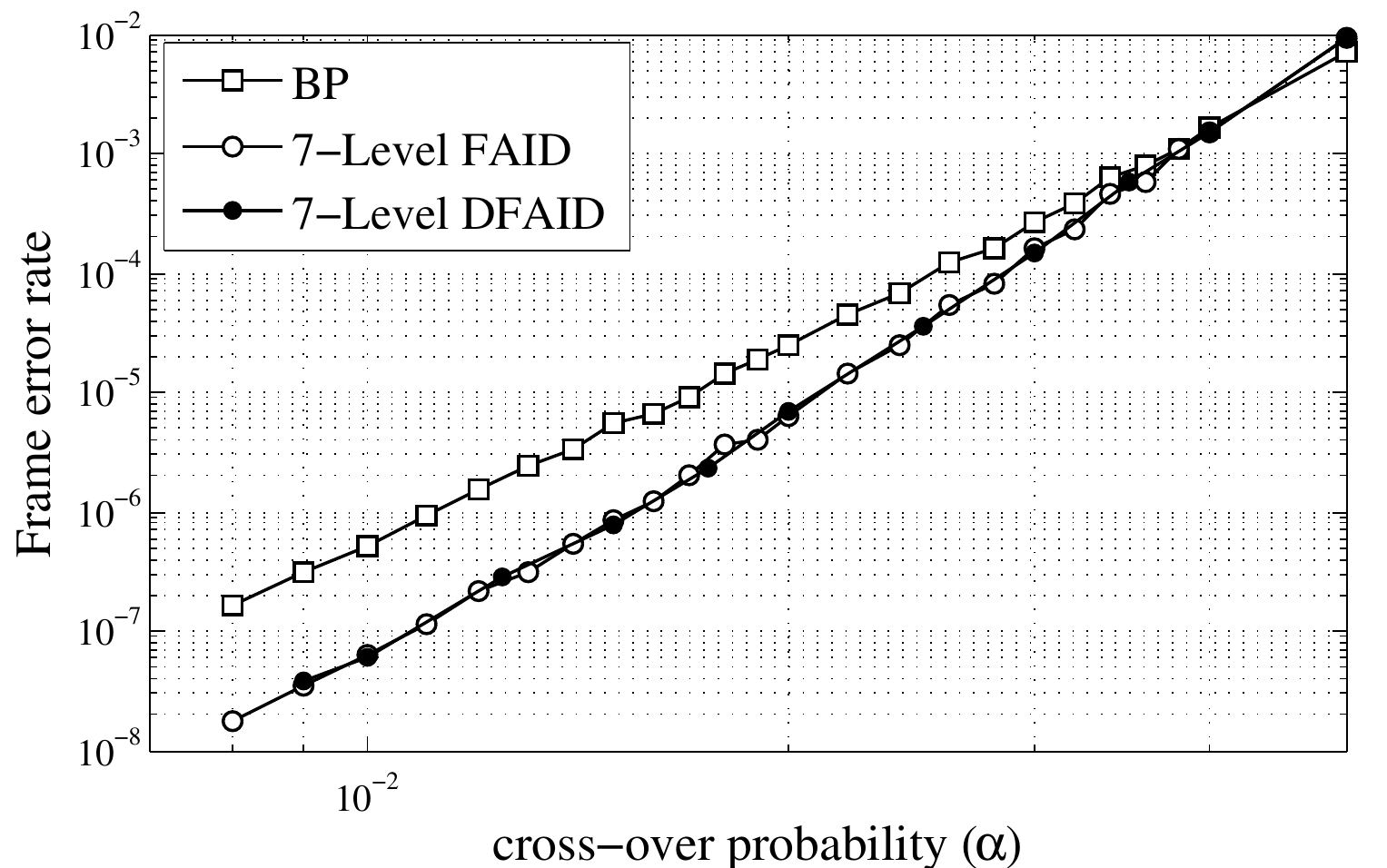}
\caption{Frame error rate performance comparison of  Belief Propagation (BP), Finite Alphabet Iterative Decoder (FAID) and Decimation-enhanced FAID (DFAID) on the (155,93) Tanner code}
\vspace{-0.2in}
\label{performance}
\end{center}
\end{figure}

%\begin{figure}[tp]
%\begin{center}
%\includegraphics[angle=0, width=2.4in]{figs/TannerPlots.pdf}
%\caption{Avg number of iterations for each decoder}
%\end{center}
%\end{figure}

%
%\begin{figure}[bh]
%\begin{center}
%\includegraphics[angle=0, width=3in]{figs/TannerPlots.pdf}
%\caption{Put only to account for room. will REPLACE}
%\label{Tanner}
%\end{center}
%\end{figure}

%
%

%
%\appendix
\section*{Acknowledgment}
This work was funded by NSF under the grants CCF-0830245 and CCF-0963726, and by the NANO2012 project.

\end{document}